\documentclass[finallayout,an,fleqn]{w-art}
\usepackage{times}
\usepackage{w-thm}
\theoremstyle{plain}

\theoremstyle{definition}

\newcommand{\3}{$_3$}
\newcommand{\kms}{km~s$^{-1}$}
\newcommand{\simgt}{\lower.5ex\hbox{$\; \buildrel > \over \sim \;$}}
\newcommand{\simlt}{\lower.5ex\hbox{$\; \buildrel < \over \sim \;$}}
\newcommand{\bq}{\begin{equation}}
\newcommand{\eq}{\end{equation}}
\usepackage[]{graphicx}
\chardef\bslash=`\\ 

\begin{document}
\DOIsuffix{theDOIsuffix}
\Volume{324}
\Issue{S1}
\Copyrightissue{S1}
\Month{01}
\Year{2003}
\pagespan{1}{}
\Receiveddate{19 February 2003}
\Reviseddate{00 March 2003}
\Accepteddate{00 March 2003}
\Dateposted{00 March 2003}
\keywords{Galactic Center, ISM, radio lines}
\subjclass[pacs]{04A25}



\title[Hot Molecular Gas]{Hot Molecular Gas in the Central 10 Parsecs of the Galaxy}


\author[R.~M. Herrnstein]{Robeson~M. Herrnstein\footnote{Corresponding
     author: e-mail: {\sf rherrnstein@cfa.harvard.edu}, Phone: 617\,495\,4142,
     Fax: 617\,496\,7554}\inst{1}} \address[\inst{1}]{MS-10, 60 Garden St., Cambridge, MA 02138}
\author[P.~T.~P. Ho]{Paul~T.~P. Ho\inst{1}}
\begin{abstract}

We present results from observations of NH\3(1,1), (2,2), (3,3), and
(6,6) with the Very Large Array.  The data sample the inner 10~pc
(4$'$) of the Galaxy and have a velocity coverage of --140 to
+130~\kms.  The velocity integrated NH\3(3,3) image shows that the Sgr
A East supernova remnant is impacting the 50~\kms ~GMC in the
northeast, the northern ridge in the north, and the western streamer
in the west.  These results imply that the Sgr A East has a large
effect on the molecular environment near Sgr A* and may be pushing
much of the molecular gas away from Sgr A*.  The physical properties
of the western streamer and its relation to Sgr A East are discussed
in detail.  We also summarize the detection of hot molecular gas less
than 2~pc from Sgr A* in projected distance.  This gas is seen only in
NH\3(6,6) and has line widths of 75--85~\kms, indicating that it is
physically close to the nucleus.

\end{abstract}
\maketitle                   





\section{Introduction}

At a distance of only $8.0\pm0.5$~kpc (Reid 1993), the Galactic Center
provides a unique opportunity to study in detail the environment
around a supermassive black hole.  It is now generally accepted that a
black hole of $\sim2.6\times10^6$M$_\odot$ is located at the dynamical
center of the Galaxy (Eckart \& Genzel 1997; Ghez et al. 1998, 2000;
Sch{\" o}del et al. 2002).  In the radio, emission from the inner region of
the accretion flow is observed as the strong ($\sim1$~Jy) source, Sgr
A*.  Sgr A* is surrounded by arcs of ionized gas (Sgr A West) that
appear to be feeding the nucleus (Lo \& Claussen 1983; Roberts \& Goss
1993).  These arcs are, in turn, surrounded by an apparent ``ring'' of
molecular material at a radius of $\sim2$~pc from Sgr A* called the
circumnuclear disk (CND, G\"{u}sten et al. 1987).  Sgr A*, Sgr A West, and
the CND appear to be located near the front edge of the expanding
supernova remnant (SNR), Sgr A East, but the exact position of the
features along the line-of-sight is very difficult to determine
(Pedlar et al 1989; Maeda et al. 2002).

For the past two decades, the origin of the clouds in the CND and the
mini-spiral has remained unclear. Many attempts have been made to
detect connections between two nearby giant molecular clouds (GMCs)
and the CND.  Okumura et al. (1989), Ho et al. (1991), and Coil \& Ho
(1999,2000) detect a long filamentary ``streamer'' in NH\3(1,1) and
(2,2) emission that connects the ``20 km s$^{-1}$ cloud''
(M--0.13--0.08; G\"{u}sten, Walmsley, \& Pauls 1981) to the
southeastern edge of the CND.  A small velocity gradient along this
``southern streamer'' as well as heating and increased line widths as
the streamer approaches the Galactic Center indicate that gas may be
flowing from the 20 km s$^{-1}$ GMC towards the circumnuclear region.
This connection has also been observed in HCN(3-2) (Marshall et
al. 1995), $^{13}$CO(2-1) (Zylka, Mezger , \& Wink 1990), and 1.1 mm
dust (Dent et al. 1993).  Other candidates for connections between the
GMCs and the CND include a possible connection between the
northeastern edge of the CND and the ``50 km s$^{-1}$ cloud''
(M-0.03--0.07; Ho 1993) and a second connection between the
20km~s$^{-1}$ GMC and the southwest lobe of the CND (Coil \& Ho
1999,2000).

Spectral line observations of the Galactic center are inherently
difficult due to the large range of velocities in the region.  In
order to detect all of the emission from the CND, a velocity coverage
of at least $\pm110$~\kms ~is necessary.  Near Sgr A*, clouds with
velocities as high as --185~\kms ~have been detected (Zhao, Goss, \&
Ho 1995).  Previous NH\3 observations by Coil \& Ho (1999,2000)
focused on the kinematics of the 20 and 50 km~s$^{-1}$ GMCs using a
velocity window of 75~\kms ~and a velocity resolution of 4.9~\kms.
Pointings were made towards Sgr A* as well as in the direction of
these GMCs in the south and east.

In order to produce a more complete picture of the molecular
environment at the Galactic center, we have imaged the central 10~pc
of the Galaxy in NH\3(1,1), (2,2), (3,3), and (6,6) with the Very
Large Array\footnote{The National Radio Astronomy Observatory is a
facility of the National Science Foundation operated under cooperative
agreement by Associated Universities, Inc.}  (VLA).  These data fully
sample the inner 10~pc (4$'$) of the Galaxy and have a velocity
coverage of --140 to +130~\kms.  In this article, we summarize two of
the most important results of this project, but we also refer the
reader to McGary, Coil, \& Ho (2001), McGary \& Ho (2002), and
Herrnstein \& Ho (2002, 2003) for more detailed discussions.  The
observations and data reduction are summarized in Section \ref{obs}.
Section \ref{snr} focuses on the velocity integrated NH\3(3,3) image,
which indicates that Sgr A East has a large effect on the molecular gas in
the region.  In Section \ref{hot}, we present the results of our
observations of NH\3(6,6) at the Galactic Center and the detection of
hot molecular gas less than 2~pc in projected distance from Sgr A*.

\section{Observations and Data Reduction\label{obs}}

The metastable (J=K) NH\3(J,K) rotation inversion transitions at
$\sim23$ GHz have proven to be useful probes of dense (10$^4$--10$^5$
cm$^{-3}$) molecular material near the Galactic center.  They tend to
have a low optical depth and a high excitation temperature at the
Galactic center, making them almost impervious to absorption effects
(Ho \& Townes 1983).  Satellite hyperfine lines separated by
10--30~\kms ~on either side of the main line enable a direct
calculation of the optical depth of NH\3, although the large line
widths at the Galactic center make it necessary to model effects due
to blending of the line profiles (Herrnstein \& Ho 2002).  In
addition, line ratios of different transitions can be used to
calculate the rotational temperature, $T_R$, of the gas.

NH\3(1,1), (2,2), and (3,3) were observed with the VLA in 1999 March.
Observations were made in the D north-C array, which provides the most
circular beam at the low elevation of the Galactic Center.  With a
maximum projected baseline of 1~km, this smallest configuration of the
VLA provides the most sensitivity to extended features such as long,
filamentary streamers.  A five-pointing mosaic was centered on Sgr A*
($\alpha_{2000}=17^h45^m40^s.0$, $\delta_{2000}=-29^\circ00'26''.6$),
with the remaining four pointings offset by $\sim1'$ to the northeast,
northwest, southeast and southwest.  The resulting data fully sample
the central 4$'$ (10~pc) of the Galaxy.  By using a velocity
resolution of $9.8$~\kms, we were also able to obtain a velocity
coverage of --140 to +130~\kms, including almost all of the velocities
observed near the nucleus.  With these data, we can probe the
morphology and kinematics of the entire CND as well as the surrounding
molecular material.

Each pointing was calibrated separately using {\it AIPS}.  (A detailed
discussion of the data reduction can be found in McGary, Coil, \& Ho
2001.)  The data were then combined in the uv-plane and deconvolved
using the Maximum Entropy Method in {\it MIRIAD}.  A Gaussian taper
was applied to the data to aid in the detection of extended features.
The final beam size for all three transitions is roughly
$15''\times13''$ with a PA of $\sim0^\circ$.  For the NH\3(3,3)
velocity integrated image, the $1\sigma$ noise level (calculated
assuming line emission typically appears in seven channels) is
$\sigma_{33}=0.33$~Jy~beam$^{-1}$~\kms ~(McGary, Coil, \& Ho 2001).

\begin{figure}[htb]
\includegraphics[width=\textwidth]{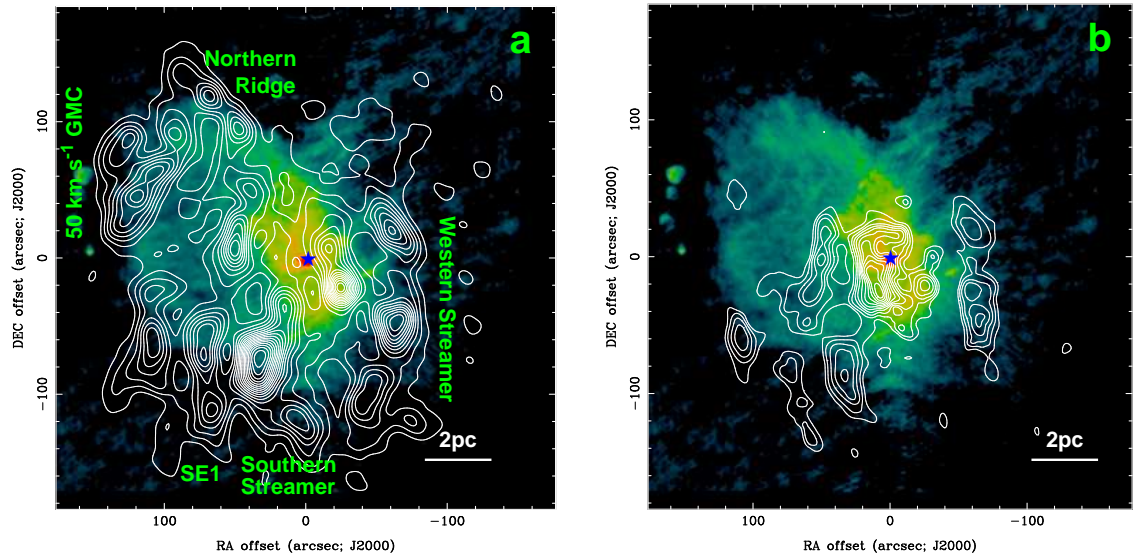}
\caption{{\bf a)} Velocity integrated NH\3(3,3) emission in contours
in steps of $4\sigma_{33}$ overlaid on a 6~cm continuum emission image
made by Yusef-Zadeh \& Morris (1987).  The position of Sgr A*
($\delta_\alpha=0'', \delta_\delta=0''$) is marked by a star.  The
northern ridge, western streamer, and 50~\kms~GMC lie along the edge
of Sgr A East (McGary, Coil, \& Ho 2001).  {\bf b)} Velocity
integrated NH\3(6,6) emission in contours in steps of $3\sigma_{66}$
overlaid on the same continuum image.  The NH\3(6,6) image is
dominated by emission within 1.5~pc ($40''$) of Sgr A*, {\it interior}
to the circumnuclear disk (Herrnstein \& Ho 2002). }
\label{mom0}
\end{figure}

Observations of NH\3(6,6) were made on 2001 October 1 and November 16
with a setup and spatial coverage identical to our previous NH\3 data.
Observations of this line, which has a frequency of 25~GHz, only
recently became possible at the VLA after the upgrade of the 23~GHz
receivers.  These data represent the first successful observation of
this line with the VLA.  The data calibration and image deconvolution
also used the same method as our original NH\3 observations.  The
final beam size after application of a Gaussian taper (FWHM=$10''$) to
the uv data is $12.0''\times9.2''$ with a PA of $-1.52^\circ$.  For
the velocity integrated image, the $1\sigma$ noise level is
$\sigma_{66}=0.23$~Jy~beam$^{-1}$~\kms.

\section{The Effect of Sgr A East on the Molecular Environment\label{snr}}

Figure \ref{mom0}a shows the velocity integrated NH\3(3,3) image in
contours overlaid on a 6~cm continuum image (Yusef-Zadeh \& Morris
1987). The contours are in steps of $4\sigma_{33}$.  In the continuum
image, the point source Sgr A* ($\Delta\alpha=0''$,
$\Delta\delta=0''$) is the brightest feature and is labeled with a
star.  The Sgr A East shell appears as faint, extended emission with a
roughly circular shape and centered slightly to the east of Sgr A*.
NH\3(3,3) is detected throughout most of the central 10~pc and the
major features discussed in this paper are labeled in Figure
\ref{mom0}a.  The gain of the telescope goes to zero at the edge of
the mosaic, roughly 5~pc from Sgr A*.  The 50 km~s$^{-1}$ GMC extends
beyond the mosaic to the northeast while the 20 km~s$^{-1}$ cloud is
located almost entirely outside our mosaic, to the south of the
southern streamer.  However, the northern ridge and western streamer
are located within the edge of our mosaic and their filamentary
morphologies are real.

\begin{figure}[htb]
\includegraphics[width=\textwidth]{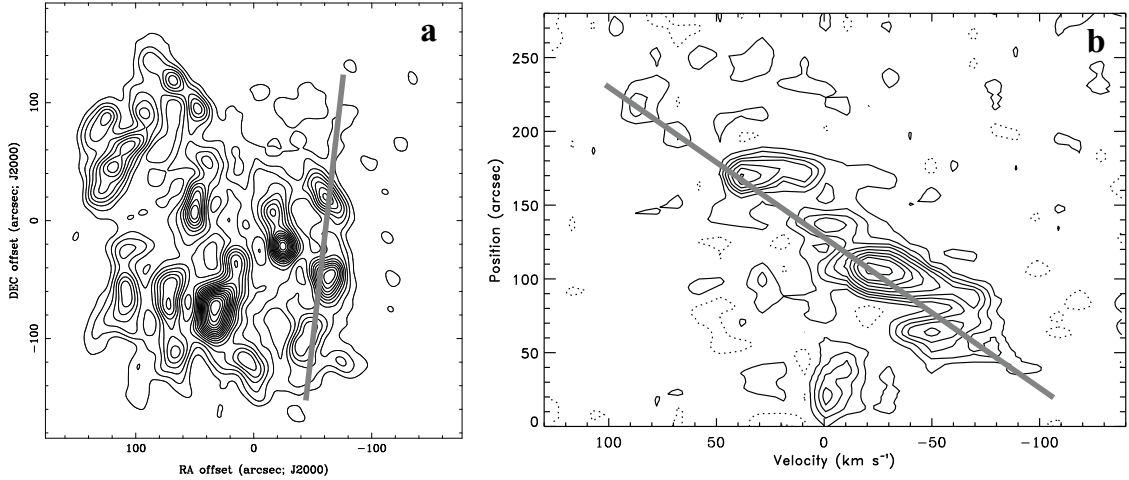}
\caption{{\bf a)} Velocity integrated NH\3(3,3) image showing the
location of the position-velocity cut through the western streamer.
{\bf b)} Position velocity diagram showing a 1~\kms ~gradient along
the entire length of the western streamer.  The $0''$ position is at
the southern end of the cut.  Emission at 0~\kms ~at a position of
$30''$ is associated with the northern edge of the 20~\kms ~GMC.}
\label{ws}
\end{figure}

Some of the clouds in our image, including SE1 and the southern
streamer, appear to be kinematically connected to gas near the nucleus
(McGary, Coil, \& Ho 2001).  However, much of the gas appears to lie
along the edge of the expanding Sgr A East supernova remnant (SNR).
In the northeast, Sgr A East is impacting the 50 km~s$^{-1}$ GMC.
Originally, it was believed that the molecular cloud had been pushed
away from the nucleus by the expanding shell (Mezger et al. 1989;
Pedlar et al. 1989).  However, with a mass of
$\sim6\times10^4$~M$_\odot$ (G\"{u}sten, Walmsley, \& Pauls 1981;
Mezger et al. 1989), it seems unlikely that Sgr A East has a large
effect on the 50 km~s$^{-1}$ GMC.  The physical properties of the NH\3 along
the edge of the 50 km~s$^{-1}$ GMC appear to support this theory.  The
intrinsic line width of gas in the 50 km~s$^{-1}$ cloud is $\sim15$~km~s$^{-1}$,
roughly equal to the mean line width throughout the inner 10~pc.  In
addition, one would expect the gas to be heated if it is being moved
by the expanding shell.  The rotational temperature of the gas can be
calculated from our NH\3(1,1) and (2,2) data by \bq
\rm{T_{R21}}=\frac{-41.5~\rm{K}}{\rm{ln}[\frac{-0.282}{\tau_m(1,1)}\rm{ln}(1-\frac{\Delta
T_A(2,2)}{\Delta T_A(1,1)}(1-e^{-\tau_m(1,1)}))]}~, \eq
\noindent where $\frac{\Delta T_A(2,2)}{\Delta T_A(1,1)}$ is the ratio
of the main hyperfine line of NH\3(2,2) and (1,1) and $\tau_m(1,1)$ is
the opacity of the NH\3(1,1) main hyperfine line (Ho \& Townes 1983).
This equation assumes equal excitation temperatures and beam filling
factors for both transitions.  For the 50 km~s$^{-1}$ cloud, we calculate a
temperature of $\sim20$~K (Herrnstein \& Ho 2003), corresponding to a
kinetic temperature of 25~K (Walmsley \& Ungerechts 1983; Danby et
al. 1988).

With a slightly elevated rotational temperature of 25~K, the northern
ridge may have been heated by the impact of Sgr A East.  However, it
is the western streamer that appears to be most strongly affected by
the expansion of the SNR.  The western streamer has a striking
velocity gradient of 1 km~s$^{-1}$ arcsec$^{-1}$ (25 km~s$^{-1}$
pc$^{-1}$) along its entire length of 150$''$ (6~pc) (See Figure
\ref{ws}).  The velocity gradient can be explained by a ridge of gas
moving outwards with the expansion of Sgr A East and highly inclined
to the line-of-sight.  This scenario would place the southern part of
the streamer on the front side of the shell.  The western streamer
also shows the largest rotational temperature of any feature in our
map, with $T_{R21}\approx50$~K, or a kinetic temperature near 80~K
(Herrnstein \& Ho 2003).  Assuming an abundance of NH\3 relative to
H$_2$ (X(NH\3)) of $10^{-7}$ (Townes et al. 1983; Harju, Walmsley, \&
Wouterloot 1993), the total mass of the western streamer is
$\sim10^2$~M$_\odot$, more than two orders of magnitude less than the
50~\kms ~GMC (Herrnstein \& Ho 2003).  Observations of hot NH\3 cores
indicate that X(NH\3) is elevated in warm environments and can be as
high as $\sim10^{-4}$ (H\"{u}ttemeister et al. 1993).  The mass
estimate for the western streamer is therefore assumed to be an upper
limit.  It is not surprising that this less massive feature shows much
more evidence for interaction with Sgr A East than the 50 km~s$^{-1}$
GMC.

\begin{figure}[htb]
\includegraphics[width=\textwidth]{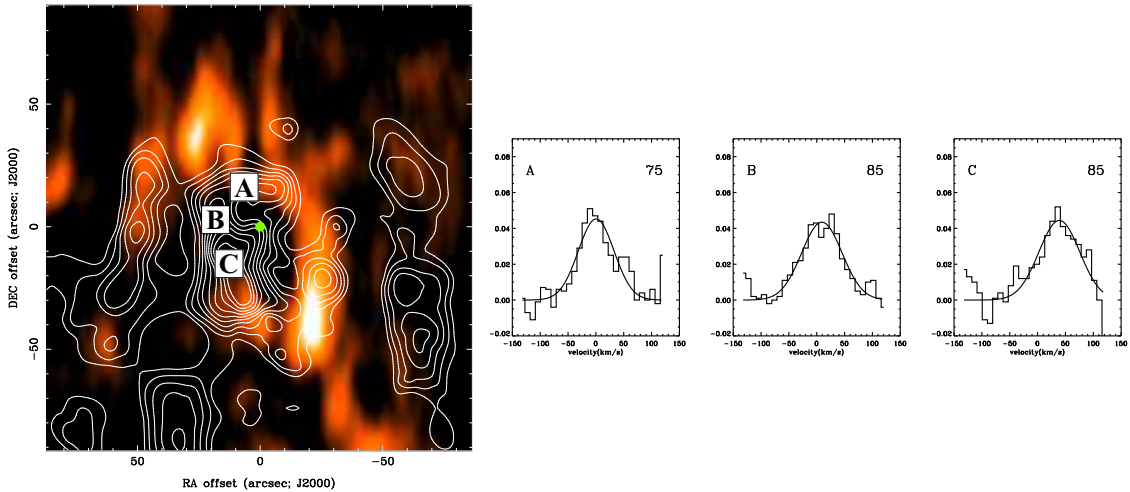}
\caption{{\bf Left:} Velocity integrated NH\3(6,6) in contours overlaid
on velocity integrated HCN(1-0) emission (Wright et al. 2001).  Sgr A*
is located at ($\delta_\alpha=0'', \delta_\delta=0''$) and is marked
by a small circle.  Much of the NH\3(6,6) is located interior to the
clumpy ring of the CND seen in the HCN(1-0) image. {\bf Right:} Spectra at the
three positions labeled in Figure \ref{cnd}a showing the large line
widths of gas interior to the CND.}
\label{cnd}
\end{figure}

In McGary, Coil, \& Ho (2001), we detected evidence for a connection
between the northern ridge and the northeastern lobe of the CND.  A
physical connection between these two features must imply that Sgr A
East is close to the CND.  The large impact of Sgr A East on molecular
gas in the central 10~pc has led us to question whether observed
connections between the GMCs and CND represent infalling material.  It
is possible that the Sgr A East shock front recently passed through
Sgr A*, and is now pushing material {\it away} from the supermassive
black hole. This scenario has also been used to explain other features
at the Galactic center including an ionized gas halo surrounding Sgr A
East (Maeda et al. 2002) and high negative velocity features seen near
Sgr A* (Yusef-Zadeh, Melia, \& Wardle 2000).  Although more data are
necessary to confirm or refute this theory, it is clear that a
detailed understanding of Sgr A East may be necessary before we can
fully understand Sgr A* and the central few parsecs of the Galaxy.

\section{Hot Molecular Gas near Sgr A*\label{hot}}

Molecular gas is expected to be heated as it approaches the
nucleus. An increased rotational temperature in the part of the
southern streamer closest to the CND has been used to argue that the
cloud is physically close to Sgr A* (Coil\& Ho 1999).  However, the
reality of this effect has remained in doubt because the emission also
becomes faint near Sgr A*.  In Figure \ref{mom0}a, the NH\3(3,3)
emission weakens and almost disappears near Sgr A*.  The lack of
emission in the inner 2~pc of the Galaxy could signal the existence of
a molecular hole near Sgr A*.  We suspected that temperatures become
so high near Sgr A* that even the NH\3(3,3) transition is no longer
well-populated.  In order to detect hot molecular gas near Sgr A*, we
observed the central 4$'$ (10~pc) of the Galaxy in NH\3(6,6) using the
new 23~GHz receivers at the VLA.  At 412~K above ground, NH\3(6,6) has
more than three times the equivalent energy of NH\3(3,3).

Figure \ref{mom0}b shows velocity integrated NH\3(6,6) emission in
contours in steps of $3\sigma_{66}$ overlaid on the same 6~cm
continuum image.  NH\3(6,6) is detected in many of the features seen
in lower NH\3 transitions, including the western streamer, but the
image is dominated by emission less than 1.5~pc (40$''$) in projected
distance from the nucleus.  Figure \ref{cnd}a overlays the velocity
integrated NH\3(6,6) image on a grey-scale image of velocity
integrated HCN(1-0) (Wright et al. 2001).  The HCN(1-0) shows the
bright clumps that form the inner ring of the CND.  Much of the
NH\3(6,6) emission comes from a region {\it interior} to the CND in
projection.

Spectra taken at three positions interior to the CND are plotted in
Figure \ref{cnd}b.  The large line widths of 75--85~\kms ~in the
central 1.5~pc indicate that this gas is physically close to the
nucleus.  Spectra are well-fitted by Gaussian profiles.  Although the
shape of the line profiles are similar to those from the CND for other
molecules (e.g. Wright et al. 2001), the emission appears to be
kinematically independent of material in the CND (Herrnstein \& Ho
2002).

\begin{figure}[htb]
\begin{center}
\includegraphics[width=.6\textwidth]{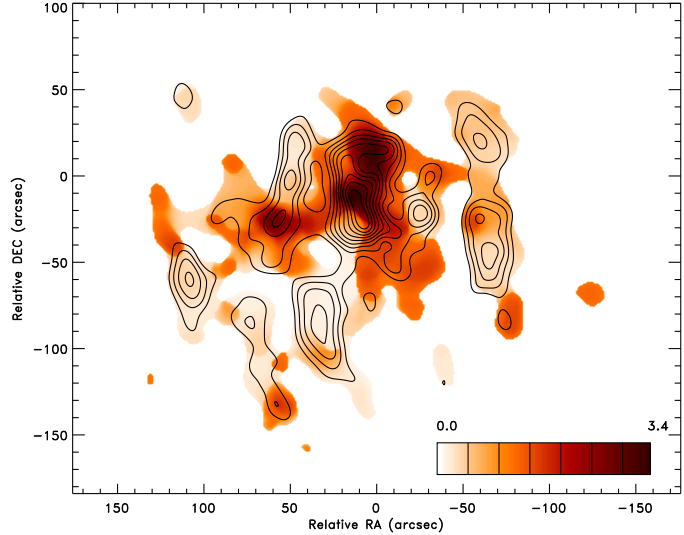}
\end{center}
\caption{Ratio of main line NH\3(6,6) and (3,3) emission in grey scale
overlaid on contours of velocity integrated NH\3(6,6). The velocity
channel for the main line is chosen using the NH\3(6,6) image cube.
Line ratios are calculated for every point with (6,6) emission
$>3\sigma_{33}$.  For pixels with faint (3,3) emission, $3\sigma_{33}$
is used to estimate a lower limit for the line ratio. Sgr A* is
located at (0,0).}
\label{lrat}
\end{figure}

The detection of NH\3(6,6) emission from a cloud is a strong
indication that the cloud is very warm.  In our data, clouds in which
NH\3(6,6) is detected tend to have high (2,2)-to-(1,1) rotational
temperatures.  Figure \ref{lrat} shows the line ratio of NH\3(6,6) to
(3,3). The NH\3(6,6) has been convolved to the resolution of the
NH\3(3,3) data.  Because NH\3(3,3) is so faint near the nucleus, we
use $3\sigma_{33}$ as the NH\3(3,3) flux density to calculate a lower
limit of the line ratio for those pixels with faint (3,3) emission.
Interior to the CND, line ratios of NH\3(6,6) to (3,3) exceed the
theoretical limit of 2.3 (Herrnstein \& Ho 2002).  These large line
ratios may the the result of a larger filling factor for NH\3(6,6), or
the dynamic range of the NH\3(3,3) data may be limited by nearby
bright emission (McGary, Coil, \& Ho 2001). It is unlikely that the
gas is out of thermal equilibrium because the equilibration time is
$\sim10^3$~s.  Line widths are quite large in the region (50 -- 80
\kms) making it unlikely for the NH\3(6,6) population to be inverted.
The large line widths also make absorption of NH\3(3,3) by an
un-associated, cool foreground cloud unlikely.  

If the NH\3(6,6) emission originates in a radiatively heated cloud,
then it is possible that the NH\3(3,3) is absorbed by cooler material
in the same cloud (with the same line width) that has been shielded
from the radiation.  The NH\3(6,6) would be unaffected by absorption
because the cooler gas would contain almost no NH\3(6,6).  This
shielded layer of cool gas must be located between the heated layer of
the cloud and the observer.  Therefore, if the clouds with line ratios
greater than 2.3 are heated by photons emanating from the nucleus,
then they must be located in front of the nucleus along the
line-of-sight.

\section{Summary}

Using the VLA, we have observed NH\3(1,1), (2,2), (3,3), and (6,6) to
investigate the physical properties of molecular gas in the central
10~pc of the Galaxy.  These data fully sample a $4'$ field centered on
Sgr A* with a velocity coverage of --140 to +130 km~s$^{-1}$.  Much of the
NH\3 emission originates along the edge of Sgr A East, indicating that
this expanding shell is greatly affecting the physical environment
near Sgr A*.  In addition, a hot component of molecular gas appears to
be located less than 2 pc from Sgr A*.  This gas has line widths of
more than 70 km~s$^{-1}$ and high (6,6)-to-(3,3) line ratios, indicating that
it is physically close to the nucleus.


\begin{thebibliography}{10}

\bibitem{coi99} Coil, A.~L. \& Ho, P.~T.~P. 1999, ApJ, {\bf 513}, 752
\bibitem{coi00} Coil, A.~L. \& Ho, P.~T.~P. 2000, ApJ, {\bf 533}, 245
\bibitem{dan88} Danby, G., Flower, D.~R., Valiron, P., Schilke, P., \& Walmsley, C.~M. 1988, MNRAS, {\bf 235}, 229
\bibitem{den93} Dent, W.~R.~F., Matthews, H.~E., Wade, R. , \& Duncan, W.~D. 1993, ApJ, {\bf 410}, 650
\bibitem{eck97} Eckart, A. \& Genzel, R. 1997, MNRAS, {\bf 284}, 576
\bibitem{ghe98} Ghez, A.~M., Klein, B.~L., Morris, M. , \& Becklin, E.~E. 1998, ApJ, {\bf 509}, 678
\bibitem{ghe00} Ghez, A.~M., Morris, M., Becklin, E.~E., Tanner, A., \& Kremenek, T. 2000, Nature, {\bf 407}, 349
\bibitem{gus87} G\"{u}sten, R., Genzel, R., Wright, M.~C.~H., Jaffe, D.~T., Stutzki, J. , \& Harris, A.~I. 1987, ApJ, 318, 124
\bibitem{gus81} G\"{u}sten, R., Walmsley, C.~M. , \& Pauls, T. 1981, A\&A, {\bf 103}, 197
\bibitem{harju93} Harju, J., Walmsley, C.~M., \& Wouterloot, J.~G.~A. 1993, A\&AS, {\bf 98}, 51
\bibitem{her02} Herrnstein, R.~M. \& Ho, P.~T.~P. 2002, ApJL, {\bf 579},83
\bibitem{her03} Herrnstein, R.~M. \& Ho, P.~T.~P., in preparation
\bibitem{ho93} Ho, P.~T.~P. 1993, in Proc. of the 2$^{nd}$ Cologne-Zermatt Symposium, The Physics and Chemistry of Interstellar Molecular Clouds, ed. G. Winnewisser \& G. Pelz (New York: Springer), 33
\bibitem{ho91} Ho, P.~T.~P., Ho, L.~C., Szczepanski, J.~C., Jackson, J.~M., Armstrong, J.~T. , \& Barrett, A.~H. 1991, Nature, {\bf 350}, 309
\bibitem{ho83} Ho, P.~T.~P. \& Townes, C.~H. 1983, ARA\&A, {\bf 21}, 239
\bibitem{hut93} H\"{u}ttemeister, S., Wilson, T.~L., Henkel, C., \& Mauersberger, R. 1993, A\&A, {\bf 276}, 445
\bibitem{lo83} Lo, K.~Y. \& Claussen, M.~J. 1983, Nature, {\bf 306}, 647
\bibitem{mae02} Maeda, Y. et al. 2002, ApJ, {\bf 570}, 671
\bibitem{mar95} Marshall, J., Lasenby, A.~N. , \& Harris, A.~I. 1995, MNRAS, {\bf 277}, 594
\bibitem{mcg01} McGary, R.~S., Coil, A.~L., \& Ho, P.~T.~P. 2001, ApJ, {\bf 559}, 326
\bibitem{mcg02} McGary, R.~S. \& Ho, P.~T.~P. 2002, ApJ, {\bf 577}, 757
\bibitem{mez89} Mezger, P.~G., Zylka, R. Salter, C.~J., Wink, J.~E., Chini, R. Kreysa, E. , \& Tuffs, R. 1989, A\&A, {\bf 209}, 337
\bibitem{oku89} Okumura, S.~K., et al. 1989, ApJ, {\bf 347}, 240
\bibitem{ped89} Pedlar, A., Anantharamaiah, K.~R., Ekers, R.~D., Goss, W.~M., van Gorkom, J.~H., Schwarz, U.~J., Zhao, J.-H. 1989, ApJ {\bf 342}, 769
\bibitem{rei93} Reid, M.~J. 1993, ARA\&A, {\bf 31}, 345
\bibitem{rob93} Roberts, D.~A. \& Goss, W.~M. 1993, ApJS, {\bf 86}, 133
\bibitem{sch02} Sch{\" o}del, R. et al. 2002, Nature, {\bf 419}, 694
\bibitem{tow83} Townes, C.~H., Genzel, R., Watson, D.~M., \& Storey, J.~W.~V. 1983, ApJL, {\bf 269}, 11
\bibitem{wal83} Walmsley, C.~M. \& Ungerechts, H. 1983, {\bf 122}, 164
\bibitem{wri01} Wright, M.~C.~H., Coil, A.~L., McGary, R.~S., Ho, P.~T.~P., \& Harris, A.~I. 2001, ApJ, {\bf 551}, 254
\bibitem{yus00} Yusef-Zadeh, T., Melia, F., Wardle, M. 2000, Science, {\bf 287}, 85
\bibitem{yus87} Yusef-Zadeh, F. \& Morris, M. 1987, ApJ, {\bf 320}, 545
\bibitem{zha95} Zhao, J.-H., Goss, W.~M., \& Ho, P.~T.~P. 1995, {\bf 450}, 122
\bibitem{zyl90} Zylka, R. Mezger, P.~G. , \& Wink, P.~E. 1990, A\&A, 234, 133

\end{thebibliography}
\end{document}